\documentclass[3p,times,twocolumn]{elsarticle}

\usepackage{ecrc}
\usepackage{slashed}

\usepackage{amsmath}

\usepackage{subfigure}
\newcommand{\wh}{\widehat}

\def\beq {\begin{equation}}
\def\eeq {\end{equation}}

\def\nn{\nonumber}

\volume{00}

\firstpage{1}

\journalname{Nuclear Physics B Proceedings Supplement}

\runauth{}

\jid{nuphbp}

\jnltitlelogo{Nuclear Physics B Proceedings Supplement}

\usepackage{amssymb}
\usepackage[figuresright]{rotating}

\def\beq {\begin{equation}}
\def\eeq {\end{equation}}
\def\bea {\begin{eqnarray}}
\def\eea {\end{eqnarray}}

\def\nn {\nonumber}

\def\HZll {H \to Z \lplm}

\def\eeHZ {e^+ e^- \to H Z}

\def\calO{\mathcal{O}}
\def\mn{\mu\nu}

\def\haV{\widehat\alpha^V_{\Phi \ell}}
\def\haA{\widehat\alpha^A_{\Phi \ell}}

\def\haVA{\widehat\alpha^{V,A}_{\Phi \ell}}
\def\ha{\widehat \alpha}

\def \wh{\widehat}

\def\bp {\mbox{\boldmath $p$}}

\def\AsymThree {\mathcal{A}_\phi^{(3)}}
\def\AsymCoCt {\mathcal{A}_{{\rm c}\theta_1,{\rm c}\theta_2}}

\newcommand{\so}{\sin\theta_1}
\newcommand{\st}{\sin\theta_2}
\newcommand{\co}{\cos\theta_1}
\newcommand{\ct}{\cos\theta_2}
\newcommand{\sosq}{\sin^2\theta_1}
\newcommand{\stsq}{\sin^2\theta_2}
\newcommand{\cosq}{\cos^2\theta_1}
\newcommand{\ctsq}{\cos^2\theta_2}

\newcommand{\sgn}{{\rm sgn}}

\def\lplm{\ell^+\ell^-}

\begin{document}

\begin{frontmatter}



%%\title{Title\tnoteref{label1}}

\title{Signatures of anomalous Higgs couplings in angular asymmetries \\of $H\to Z \ell^+ \ell^-$ and $e^+e^- \to H Z$ }
 \tnotetext[label1]{TUM preprint HEP-968/14}
  
\author[tum]{M.~Beneke}
\author[tum,usp]{D.~Boito\corref{cor1}}
\cortext[cor1]{Speaker}
\author[tum,aachen]{Y.-M.~Wang}
% \ead{diogo.boito@tum.de,}

 \address[tum]{Physik Department T31, Technische Universit\"at M\"unchen\\
James-Frack-Stra\ss e 1, D-85748 Garching, Germany}
\address[usp]{Instituto de F\'isica, Universidade de S\~ao Paulo, \\ Rua do Mat\~ao Travessa R, 187, 05508-090,  S\~ao Paulo, SP, Brazil}
\address[aachen]{Institut f\"ur Theoretische Teilchenphysik und Kosmologie, RWTH Aachen
University,\\ D-52056 Aachen, Germany}

% \ead{diogo.boito@tum.de,}
 %\fntext[label2]{abc}
 %\cortext[cor1]{def}

% \fntext[label3]{ert}



\begin{abstract}
\noindent Parametrizing beyond Standard Model physics by the $SU(3)\times SU(2)_L \times U(1)_Y$ dimension-six effective lagrangian, 
we  study the impact of anomalous Higgs couplings in angular asymmetries of the crossing symmetric processes $H \to Z \ell^+ \ell^-$ and $e^+ e^- \to H Z$.  
In the light of present bounds on $d=6$ couplings, we show that some asymmetries can reveal BSM effects that would otherwise be hidden in other observables.
The $d=6$ $HZ\gamma$ couplings as well as (to a lesser extent)  $HZ\ell\ell$ contact interactions  can generate asymmetries at the several percent level, albeit having less significant effects on the di-lepton 
invariant mass spectrum of the decay $H \to Z \ell^+ \ell^-$.
The higher di-lepton invariant mass probed in $e^+e^- \to H Z$ can lead 
to complementary anomalous coupling searches at  $e^+e^-$ colliders.
 \end{abstract}


\end{frontmatter}

% \linenumbers

\section{Introduction, operators and couplings}
\label{sec:Intro}

The discovery of a light boson $H$ with mass around 125~GeV at the LHC~\cite{Atlas_Higgs,CMS_Higgs} has opened a  window
to a new sector in the search for physics beyond the Standard Model (BSM).  The new state is  compatible with a  SM Higgs, with the quantum numbers $J^P=0^+$ being highly  favoured by
the data~\cite{Atlas_HiggsQN,CMS_HiggsQN}. The study of signal strengths of the new state has shown that the Higgs couplings are compatible with SM predictions. Evidence for
BSM physics has proven to be more elusive than previously expected; the SM appears to be a good effective field theory (EFT) at the least up to the energies probed
by the first run of LHC.

In the spirit of an EFT, the SM should be supplemented with all operators with dimension $d>4$ constructed from its fields and compatible with its symmetry. In this work we adopt the linear
realization of the $SU(2)_L\times U(1)_Y$ symmetry~\cite{BW86,GIMR10}. The leading corrections to Higgs physics within this
scheme arise from the dimension-six operators, that are suppressed by the large scale $\Lambda$ characteristic of BSM physics, and generate anomalous Higgs boson couplings.

In the search for BSM physics in the flavour sector of the standard model, in particular in the case of flavour-changing neutral currents, dedicated observables were
constructed from the angular distribution of the decay $B\to K^* \ell \ell$. The angular distribution of the decay $H\to Z \ell^+ \ell^-$ offers similar possibilities that
we exploit in this work.

The study of the decay  $H\to Z \ell^+ \ell^-$, with the on-shell $Z$ also decaying into $\ell^+\ell^-$, has a long history. Its angular distributions
were instrumental in the determination of the Higgs quantum numbers~\cite{Atlas_HiggsQN,CMS_HiggsQN}, as suggested long ago (see
e.g.~Refs.~\cite{CMMZ2003,GMM07,DeRujula_etal10,Bea12}). After the Higgs discovery, it has been suggested that the di-lepton mass distribution of the decay 
can reveal effects that would be hidden in the total decay width~\cite{IMT13,IT13,GMP13,GI14}. Recently, the full angular distribution  of the decay has been revisited
in the framework of the EFT parametrization of BSM physics~\cite{BCD13}. It has been shown that some angular asymmetries can reveal effects that would
be hidden even in the di-lepton mass distributions.

More recently, we performed an extended study of the angular asymmetries of $H\to Z(\to \lplm) \lplm$  and of the crossing-symmetric
process $e^+e^-\to H Z$~\cite{BBW14}. The latter process should be measured with precision at a high-energy $e^+e^-$ collider such as the ILC~\cite{Baeretal13}
and should provide a clean way to extract Higgs couplings~\cite{Barger:1993wt,Hagiwara:1993sw,KKZ96}.
Our focus is on these asymmetries,
their sensitivity to anomalous Higgs couplings, and the interplay
between the asymmetries and the di-lepton mass distributions. 
Our aim here is to  highlight some of the main findings of Ref.~\cite{BBW14};  for additional
details we refer to  that reference.

In the massless lepton limit,  the two processes are
described by the same set of six form factors~\cite{CMMZ2003,GMM07,IMT13,GMP13,BCD13}, albeit in different
kinematic regimes, related by analyticity.    Ignoring loop corrections and neglecting the lepton masses, 
the processes are governed  by six independent angular functions of 
 three independent angles among the four leptons. These functions can be expressed 
in terms of the six form factors that, in turn, can be
written in terms of the couplings of the general $d=6$
Lagrangian.

% \section{Effective lagrangian and couplings}

Assuming the new physics sector to be characterized by the 
scale $\Lambda$, larger than the electroweak scale, the
SM is supplemented with 59 independent $d=6$ operators~\cite{BW86,GIMR10}. 
This Lagrangian can be schematically cast~as
\beq
\mathcal{L}_{\rm eff} = \mathcal{L}_{\rm SM}^{(4)} + 
\frac{1}{\Lambda^2} \sum_{k=1}^{59} \alpha_{k}\calO_k,
\label{leff}
\eeq
where the $\alpha_k$ is the coupling of operator $\calO_{k}$. 
The effective Lagrangian implies a parametrization of anomalous 
Higgs interactions (contained in $\calO_k$) constrained by the SM gauge 
symmetry.  In our expressions, we often employ the 
dimensionless coefficients $\widehat\alpha_k$ defined as
\beq
\widehat\alpha_k = \frac{v^2}{\Lambda^2} \alpha_k,
\label{ahat}
\eeq
where  $v$ is the classical Higgs vacuum expectation value.
The  coefficients $\widehat\alpha_k$ should be smaller 
than $\mathcal{O}(1)$ for the EFT description to be applicable.
 
 Here we 
employ the complete operator basis  defined in Ref.~\cite{GIMR10}, although
different choices  are possible and in use. In practice we only need to work with a  subset of the 59
operators, since not all of them contribute at tree level to the
processes of interest. Furthermore, assuming
minimal-flavour violation to avoid tree-level flavour-changing neutral
currents, flavour matrices of operators that involve a left-handed
doublet and a right-handed singlet are fixed to be the same as in the
SM Yukawa couplings. With this assumption, these operators are
proportional to lepton masses and can be safely neglected.
  
  The operators considered in this work are listed in Eqs.~(\ref{eq:Ops1}),~(\ref{eq:Ops2}) and~(\ref{eq:Ops3}) below. The
notation and conventions follow those of Ref.~\cite{Heinemeyer_etal13} to which we refer for further details. 
The first two operators involve four Higgs doublets  and are
 \begin{eqnarray}
 \calO_{\Phi\Box}&=&(\Phi^\dagger\Phi)\Box(\Phi^\dagger \Phi), \nn \\
 \calO_{\Phi D}&=&(\Phi^\dagger D^\mu \Phi)^*(\Phi^\dagger D_\mu \Phi) \label{eq:Ops1}.
 \end{eqnarray}
 They modify the Higgs-gauge couplings and entail a redefinition of the Higgs field to preserve canonically 
 normalized kinetic terms. Operators of the form $X^2\Phi^2$, where $X=W^I,B$,  generate anomalous couplings
 of the Higgs to $ZZ$, $\gamma Z$, and $WW$. They are, explicitly, 
 \begin{eqnarray}
\calO_{\Phi W}&=&(\Phi^\dagger\Phi) W^{I}_{\mn} W^{I\mn},\nn \\ 
\calO_{\Phi B} &=& (\Phi^\dagger \Phi)B_{\mn}B^{\mn}, \nn \\ 
\calO_{\Phi W\! B} &=&(\Phi^\dagger \tau^I\Phi) W^{I}_{\mn} B^{\mn},\nn \\ 
\calO_{\Phi \widetilde W} &=&(\Phi^\dagger\Phi) \widetilde W^{I}_{\mn} W^{I\mn},\nn \\ 
\calO_{\Phi \widetilde B} &=&(\Phi^\dagger\Phi) \widetilde B_{\mn}B^{\mn},\nn \\ 
\calO_{\Phi \widetilde W\! B} &=&(\Phi^\dagger\tau ^I\Phi) \widetilde W_{\mn}^IB^{\mn}.\label{eq:Ops2}
 \end{eqnarray}
 Finally, three operators involving two fermion fields, that yield contact $HZ\ell\ell$ interactions as well as modifications
 to gauge-boson couplings to leptons, should be taken into account:
 \bea
 \calO^{(1)}_{\Phi\, \ell}&=&(\Phi^\dagger i 
\overset{\leftrightarrow}{D}_\mu \Phi)(\bar\ell\gamma^\mu \ell),\nn \\ 
\calO^{(3)}_{\Phi\, \ell}&=&(\Phi^\dagger i 
\overset{\leftrightarrow}{D} \ \!^{I}_\mu \Phi)
(\bar\ell\gamma^\mu\tau^I \ell),\nn \\ 
\calO_{\Phi e}&=&(\Phi^\dagger i 
\overset{\leftrightarrow}{D}_\mu \Phi)(\bar e \gamma^\mu e).\label{eq:Ops3}
 \eea
 
 As input parameters  we employ $G_F$ (the Fermi 
constant as measured in $\mu \to e \nu_\mu \bar \nu_e$ decay), 
the $Z$ mass $m_Z$, the electromagnetic coupling $\alpha_{\rm em}$, and 
the Higgs mass $m_H$. We trade the Lagragian parameters $g$, $g'$, the Higgs self-coupling $\lambda$, and 
the classical Higgs vacuum expectation value $v$ for combinations 
of the former.  Dimension six corrections to our input parameters must be taken into account and are discussed in detail in~\cite{BBW14}. In particular,
a four-fermion operator not listed above contributes to the redefinition of $G_F$ and must be considered~\cite{AJMT13}.

Apart from SM contributions we include a single insertion of a dim-6 operator. We neglect  $1/\Lambda^4$ terms in the square of the amplitudes (except for photon-pole enhanced terms, see Ref.~\cite{BBW14}). 
The effects of the operators listed above can be summarized in the following effective Lagragian 
\begin{align}
&\mathcal{L_{\rm eff}} \supset c_{ZZ}^{(1)} \, H Z_\mu Z^\mu + 
c_{ZZ}^{(2)} H\, Z_{\mn}Z^{\mn} + 
c_{Z\widetilde Z}H\, Z_{\mn}\widetilde Z^{\mn}\nn \\&+ 
c_{AZ}H \, Z_{\mn} A^{\mn} + 
c_{A\widetilde Z} H Z_{\mn}\widetilde A^{\mn} \nn 
+ \,H Z_\mu \bar \ell \gamma^\mu \left( c_V + c_A\gamma_5   \right) \ell  \\
&+Z_\mu \bar \ell \gamma^\mu (g_V - g_A \gamma_5)\ell -
g_{\rm em}Q_\ell A_\mu \bar \ell \gamma^\mu \ell.
\label{eq:effLag}
\end{align}
The effective Lagrangian of $HZZ$ interaction depends on the
basis of  $d=6$ operators; the above Lagrangian  is constructed from  the  complete and non-redundant
operator basis of Ref.~\cite{GIMR10}.

 The effective couplings  can be written in terms of the underlying dimension-six operators. We list here explicitly only
 three of them, namely the contact $HZ\ell\ell$ couplings $c_{V,A}$ and the CP-even anomalous $HZ\gamma$ coupling, since they enter the phenomenology discussed in the sequel. We have
 \bea
  c_V &=& \sqrt{2}G_F\,m_Z \,\widehat\alpha^V_{\Phi \ell}, \nn\\
c_A &=& \sqrt{2}G_F \,m_Z\,\widehat\alpha^A_{\Phi \ell}, \nn \\
c_{AZ} &=&  (\sqrt{2}G_F)^{1/2}\,\widehat \alpha_{AZ},\label{eq:cs}
 \eea
 with
 \begin{align}
 &\haV  = \widehat\alpha_{\Phi e} +
\left(\widehat\alpha_{\Phi \ell}^{(1)}+
\widehat\alpha_{\Phi \ell}^{(3)}\right),\nn\\
&\haA   = \widehat\alpha_{\Phi e} -  
\left(\widehat\alpha_{\Phi \ell}^{(1)}+
\widehat\alpha_{\Phi \ell}^{(3)}\right),\nn\\
 & \ha_{AZ} = 2s_W\,c_W ( \ha_{\Phi W}-\ha_{\Phi B}) + (s_W^2-c_W^2)\ha_{\Phi W\!B},\label{eq:alphas}
 \end{align}
 where $s_W$ and $c_W$ are respectively $\sin \theta_W$ and $\cos\theta_W$.
The remaining couplings are given in detail in~\cite{BBW14}.

In the study of the phenomenological  impact of the anomalous couplings one must take
into account the constraints imposed on these couplings by the present LHC data, as well as
EW precision data. We estimate~\cite{BBW14} the following bounds 
for the contact $HZ\ell\ell$ couplings 
\beq
\haVA \in [-5,5] \times 10^{-3}.
\label{contactrange}
\eeq
Also, from the results of  Ref.~\cite{PR13}, one can deduce the following bounds within our conventions
\beq
\ha_{AZ} \in [ -1.3,2.6] \times 10^{-2}.
\label{hzarange}
\eeq
We allow the above couplings to vary within these limits.

%%%%%%%%%%%%
%%%%%%%%%%%%
%%%%%%%%%%%%
%%%%%%%%%%%%
%%%%%%%%%%%%
%%%%%%%%%%%%

\section{Angular asymmetries of $H\to Z(\to \lplm) \lplm$ and $e^+e^- \to H Z(\to \ell^+ \ell^-)$}

We discuss here the angular structures of the differential decay amplitude of $H\to Z(\to \lplm) \lplm$ 
and of the cross-section of $e^+e^- \to H Z(\to \ell^+ \ell^-)$. Here we discuss in some detail the decay case; the 
description of the scattering can be done in close analogy exploiting crossing symmetry.

Summing over spins of the
final-state leptons, the four-fold differential decay width for the
process $H(p_H) \to Z(p) (\to \ell^{-}(p_1) \ell^{+}(p_2)) \ell^-(p_3) \ell^+(p_4)$ in the massless lepton limit can be written as a
function of the di-lepton invariant mass squared $q^2 = (p_3+p_4)^2$
and of three angles. We chose the angles $\theta_{1,2}$, the angles between the direction of $\bp_1$ and $\bp_3$ and the $z$-axis in the respective di-lepton rest frames, and the
angle $\phi$ between the normals of the di-lepton decay planes.   The expression for the differential decay width reads
\beq
\frac{d^4 \Gamma }{  dq^2 d \cos \theta_1 d \cos \theta_2 d \phi}= \frac{1}{m_H}\, \mathcal{N} (q^2)\,\mathcal{J}(q^2, \theta_1, \theta_2, \phi),
\label{eq:DiffDecayRate}
\eeq
with the normalization factor
\begin{eqnarray}
\mathcal{N}(q^2)=\,\frac{1 }{ 2^{10} (2 \pi)^5}  \frac{1}{\sqrt{r}\, \gamma_Z}
\lambda^{1/2}(1, r,s),\label{eq:Norm}
\end{eqnarray}
written in terms of the  dimensionless variables
\beq
s= \frac{q^2}{m_H^2},\,\, r= \frac{m_Z^2}{m_H^2}\approx 0.53, \,\, 
\,\, \gamma_Z = \frac{\Gamma_Z}{m_H}\approx 0.020,
\label{eq:AdVar}
\eeq
and  the function $\lambda(a,b,c) = a^2 +b^2 +c^2 -2ab -2ac -2bc$.
The maximum value of $q^2 $ is
$q_{\rm max}^2 = (m_H-m_Z)^2 \approx (34.4~\mbox{GeV})^2$ which gives
\beq
0\le  s \le \frac{(m_H-m_Z)^2}{m_H^2} \approx 0.075.
\eeq
The function $\mathcal{J}(q^2, \theta_1, \theta_2, \phi)$ has 
nine independent angular structures with coefficient 
functions  $J_1$,...,$J_{9}$, which we write 
\begin{eqnarray}
&&\mathcal{J}(q^2, \theta_1, \theta_2, \phi) \nn\\ 
&&=J_1(1+ \cosq\ctsq {+\cosq +\ctsq}) \nn \\ 
&& +\, J_2 \sosq\stsq +J_3\co\ct  \nn \\
&& +\, \left(J_4 \so\st +J_5\sin2\theta_1\sin2\theta_2\right)\sin\phi
\nn \\
&& +\, \left(J_6 \so\st +J_7\sin2\theta_1\sin2\theta_2\right)\cos\phi
\nn \\
&& +\,J_8\sin^2\theta_1\,\sin^2\theta_2\,\sin2\phi
\nn \\
&& 
+J_9\sin^2\theta_1\,\sin^2\theta_2\,\cos2\phi  .
\label{eq:FullJ}
\end{eqnarray}
The $J$ functions are in turn written in terms of six form-factors that we denote $H_i^{V,A}$ (with $i=1,2,3$)
\begin{align}
&J_1=2 \, r \,s\, \left(g_A^2+g_V^2\right)
\left(|H_{1,V}|^{2}+|H_{1,A}|^{2}\right), \nn \\
&J_2=\kappa\, \left(g_A^2+g_V^2\right) \big[\kappa \,
\left(|H_{1,V}|^{2 }+|H_{1,A}|^{2}\right)
\nn \\ 
&\, \, \, \,  +\lambda{\rm Re} \left( H_{1,V} H^{\ast}_{2,V} + H_{1,A} H^{\ast}_{2,A}\right)
\big],\nn \\
&J_3=32 \,r\, s\, g_A\, g_V\, {\rm Re} \left ( H_{1,V} \, H_{1,A}^{\ast} 
\right ), \nn 
\\ %\end{align}
%\begin{align}
&J_4=4\kappa\, \sqrt{r\,s\,\lambda} \,g_A\,g_V \, {\rm Re}
\left(H_{1,V} H_{3,A}^{\ast}+H_{1,A}H_{3,V}^{\ast}\right),\nn \\
&J_5=\frac{1}{2}\kappa\, \sqrt{r\,s\,\lambda} \,
\left(g_A^2+g_V^2\right) {\rm Re} \left(H_{1,V} H_{3,V}^{\ast}+H_{1,A}\,
H_{3,A}^{\ast}\right),\nn \\
&J_6=4 \sqrt{r\,s}\, g_A\, g_V\, \big[4 \kappa  \, {\rm Re}
\left (H_{1,V}H_{1,A}^{\ast} \right)\nn \\ 
& \,\,\,\, +\lambda  \, {\rm Re} \left (H_{1,V}H_{2,A}^{\ast} + H_{1,A}H_{2,V}^{\ast} 
\right) \big] ,\nn \\
&J_7=\frac{1}{2} \sqrt{r\, s} \left(g_A^2+g_V^2\right) \big[2 \kappa \, 
\left(|H_{1,V}|^{2}+|H_{1,A}|^{2}\right) \nn \\
&\,\,\,\,\, + \lambda \, {\rm Re} \left(H_{1,V} H_{2,V}^{\ast} + H_{1,A} H_{2,A}^{\ast} 
\right)
\big] ,\nn \\
&J_8=2\, r\, s\, \sqrt{\lambda } \left(g_A^2+g_V^2\right) {\rm Re}
\left( H_{1,V}H_{3,V}^{\ast}+ H_{1,A}H_{3,A}^{\ast}\right),\nn \\[0.1cm]
&J_9=2 \,r \,s \,\left(g_A^2+g_V^2\right)
\left(|H_{1,V}|^{2}+|H_{1,A}|^{2}\right),\label{eq:JFuncsinH}
\end{align}
where $\kappa=1-r-s$.
The form factors parametrize the amplitude ${\cal M}_{HZ\ell\ell}^\mu$ for the decay $H\to Z \ell\ell$
as follows
\begin{align}
&{\cal M}_{HZ\ell\ell}^\mu = \frac{1}{m_H} \,
\bar u(p_3,s_3) \bigg[ \gamma^\mu\left(H_{1,V}+H_{1,A}\,\gamma_5\right)\nn\\
& + 
\frac{ q^\mu \slashed{p}}{m_H^2} \left(H_{2,V}+H_{2,A}\,\gamma_5\right)  
\nonumber \\
&+\, \frac{\epsilon^{\mu\nu\sigma\rho}p_\nu q_\sigma}{m_H^2} \, 
\gamma_\rho \,\left(H_{3,V}+H_{3,A}\,\gamma_5\right)\bigg] 
v(p_4,s_4),\label{eq:MHZll}
\end{align}
where $\epsilon_{0123}=+1$ and $q=p_3+p_4$. 
 An important observation is that SM tree level contributions appear solely in
$H_1^{V,A}$, the form factors $H_{2,3}^{V,A}$ are suppressed by $1/\Lambda^2$. Therefore, we neglect 
higher powers of $H_{2,3}^{V,A}$ in Eq.~(\ref{eq:JFuncsinH}).

The cross section for $\eeHZ$ is governed by the same set of form factors, in another kinematic
regime and related by analyticity to those of $\HZll$. We can cast the cross section as\footnote{For the
precise definition of the three angles in this case we refer to the appendix A.2 of Ref.~\cite{BBW14}.}
\beq
\frac{d\sigma}{ d\cos\theta_1 \,d\cos\theta_2  d\phi} = \frac{1}{m_H^2} \, 
\mathcal{N_\sigma}(q^2) \, \mathcal{J}(q^2,\theta_1,\theta_2,\phi) ,
\eeq
where the new normalisation reads
\beq
\mathcal{N_\sigma}(q^2) = \frac{1}{2^{10}(2\pi)^3}\frac{1}{\sqrt{r}\, 
\gamma_Z}\frac{\sqrt{\lambda(1,s,r)}}{s^2}.
\eeq
The threshold energy for the reaction is given by $\sqrt{q_{\rm th}^2}
= (m_H+m_Z)\approx 217$~GeV which gives, in units  of $m_H^2$, the 
minimal $s$ value
\beq
s_{\rm th}= q_{\rm th}^2/m_H^2 \approx 2.98.
\eeq
The form factors are therefore probed at much higher energies, which 
leads to non-trivial phenomenological consequences in comparison 
with $\HZll$.

Integrating over the three angles, the differential decay rate and the total cross section are proportional to the
combination $4J_1 +J_2$. Angular asymmetries can be constructed to give us  access to the information
contained in other $J$ functions. Here we show results for two of them, written explicitly for the case of the decay $\HZll$. Following
the notation of Ref.~\cite{BBW14} we write
\begin{align}
&\AsymThree = \frac{1}{d\Gamma /dq^2} \, \int_{0}^{2 \pi}    
d \phi  \,\, {\rm sgn}(\cos \phi) \,
\frac{d^2 \Gamma}{d q^2 d \phi} \nn \\ 
&= \frac{9 \, \pi}{32} \,  
\frac{J_6}{4J_1+J_2} ,  \label{eq:Asm7}\\
&\AsymCoCt =\frac{1}{d\Gamma /dq^2} \nn \\ &\times \int_{-1}^{1} 
d\cos\theta_1\, \sgn(\cos \theta_1)\int_{-1}^{1} 
d\cos\theta_2\, \sgn(\cos \theta_2) \nn \\ &\qquad 
\frac{d^3 \Gamma}{  dq^2 d \cos \theta_1 d \cos \theta_2 }= \frac{9}{16} \, \frac{J_{3}}{4J_1+J_2}. \label{eq:Asm9}
\end{align}
The sign function is $\sgn(\pm |x|) = \pm 1$. The asymmetries for $\eeHZ$ can be written in close analogy to the ones above.

The functions $J_3$ and $J_6$ are sensitive to the contact $HZ\ell\ell$ interaction and to CP-even anomalous $HZ\gamma$ coupling, whose expressions are given 
in  Eqs.~(\ref{eq:cs}) and~(\ref{eq:alphas}). In the remainder we discuss briefly  how these anomalous couplings impact the asymmetries in two specific cases.
A thorough discussion of the asymmetries and couplings in both $\HZll$ and $\eeHZ$ is found in Ref.~\cite{BBW14}.

\begin{figure}[!t]
\begin{center}
\subfigure[$-\AsymThree$]{\includegraphics[width=.9\columnwidth,angle=0]{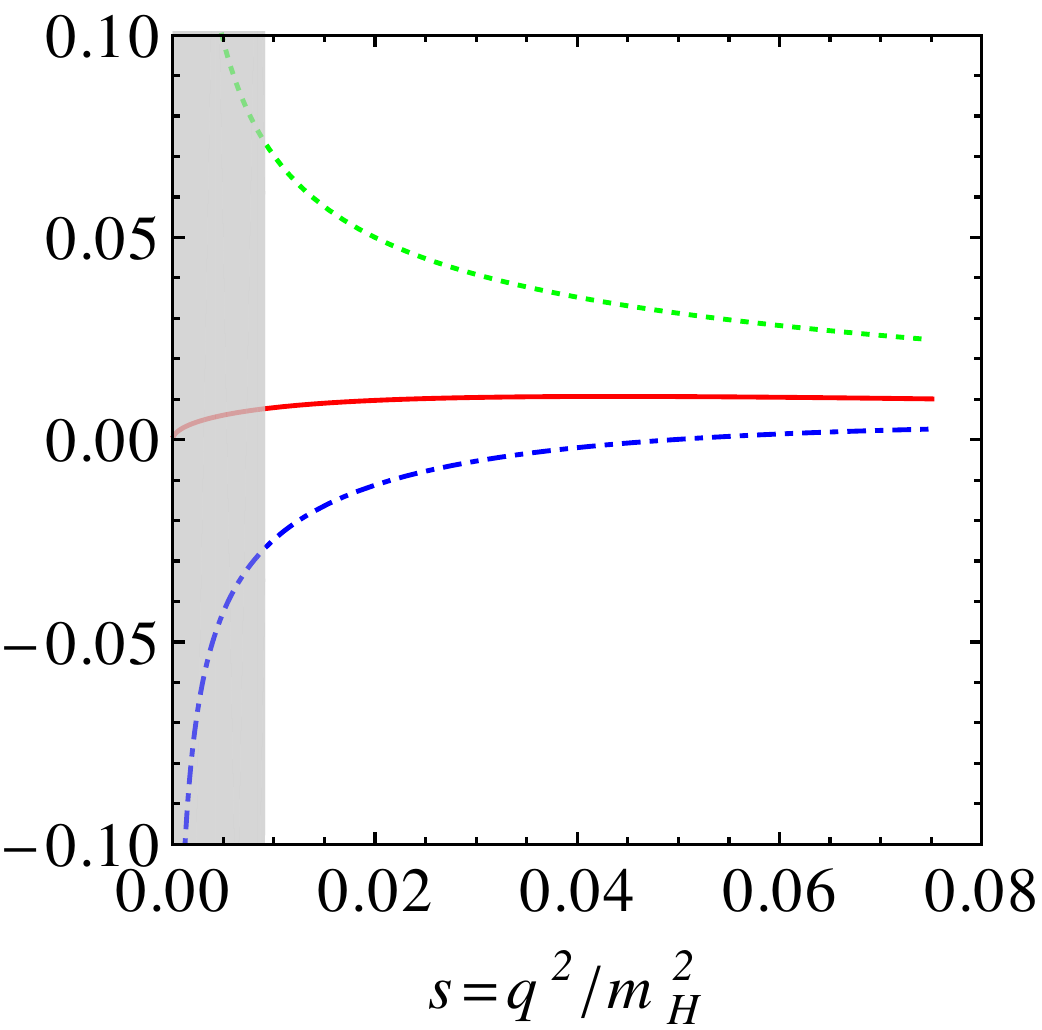}\label{Fig1a}}
\subfigure[$-\AsymCoCt$]{\includegraphics[width=.9\columnwidth,angle=0]{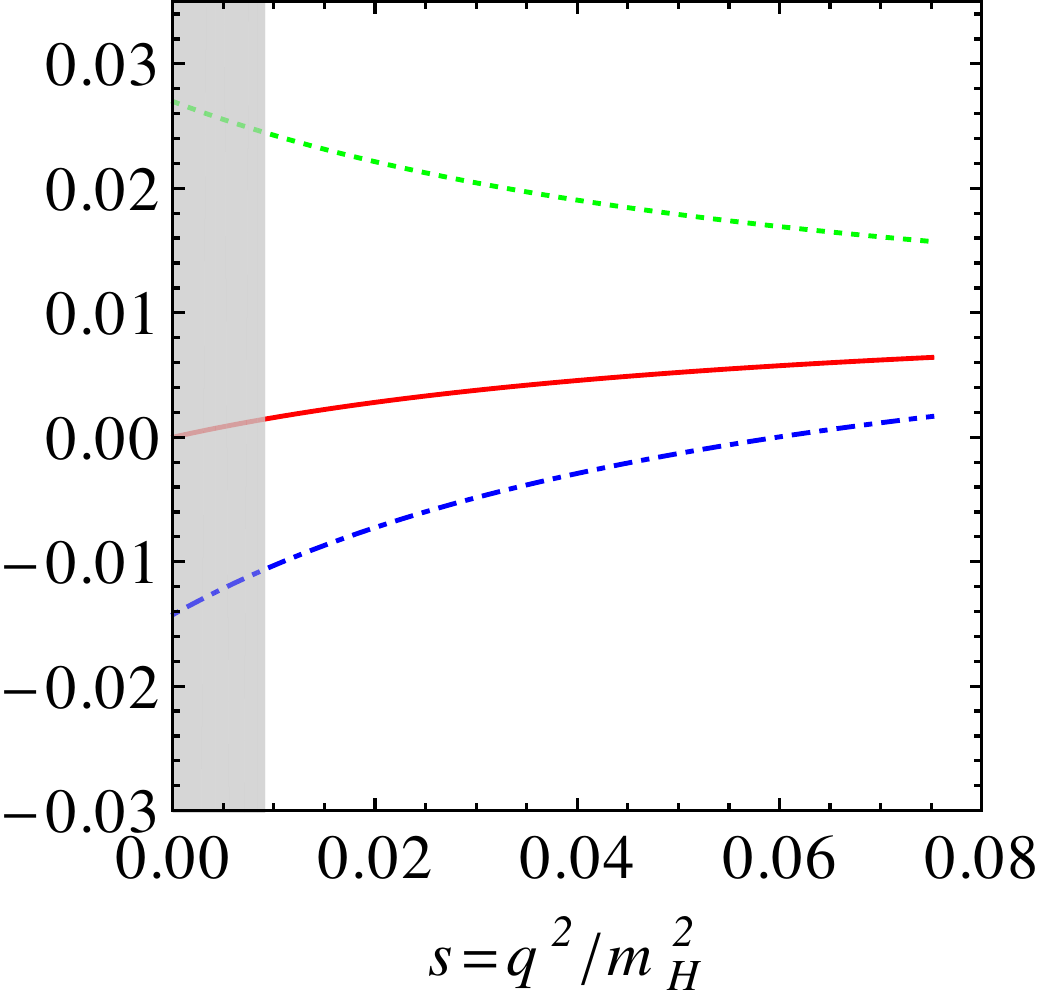}\label{Fig1b}}
\caption{ (a)  $-\AsymThree$, (b) $-\AsymCoCt$.
Three scenarios are considered. The red solid-line is the SM case. 
The dot-dashed blue line corresponds to $\wh\alpha_{AZ}=-1.3\times 10^{-2}$,
 whereas the  dotted green line corresponds to 
$\wh\alpha_{AZ}=2.6\times 10^{-2}$. The gray band excludes the region  $0\leq q^2 \leq (12\, \, \mbox{GeV})^2$ where the decay $(Z^*,\gamma^*)\to \lplm$ is 
dominated  by hadronic resonances.}\vspace{-0.8cm}
\label{fig:AsymHZll}
\end{center}     
\end{figure}

First, let us study the  asymmetries of Eqs.~(\ref{eq:Asm7}) and~(\ref{eq:Asm9})   in a scenario where we consider the impact of $\ha_{AZ}$; all other anomalous couplings are set to zero for the moment. 
The results of this exercise are shown in Fig.~\ref{fig:AsymHZll}.  The asymmetries are particularly sensitive to this anomalous coupling, in Fig.~\ref{Fig1a},~$|\AsymThree|$   reaches $\sim\!10\%$  for lower values of 
$q^2$, compared to an almost zero asymmetry in the SM. This relatively high sensitivity is, in the case of $\AsymThree$, due to the photon pole.

\begin{figure}[!t]
\begin{center}
\subfigure[$\sigma$]{\includegraphics[width=.82\columnwidth,angle=0]{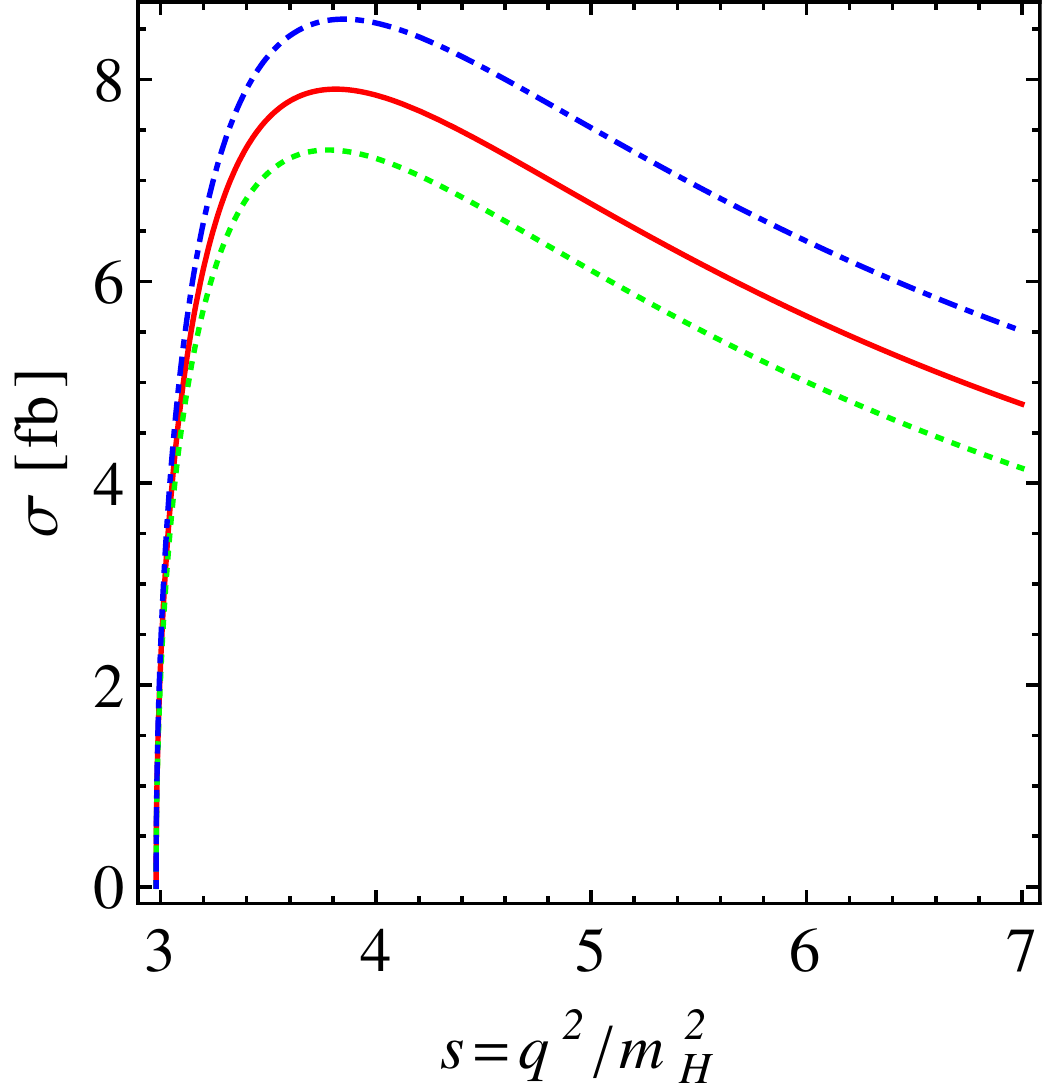}\label{Fig2a}}
\subfigure[$-\AsymThree$]{\includegraphics[width=.9\columnwidth,angle=0]{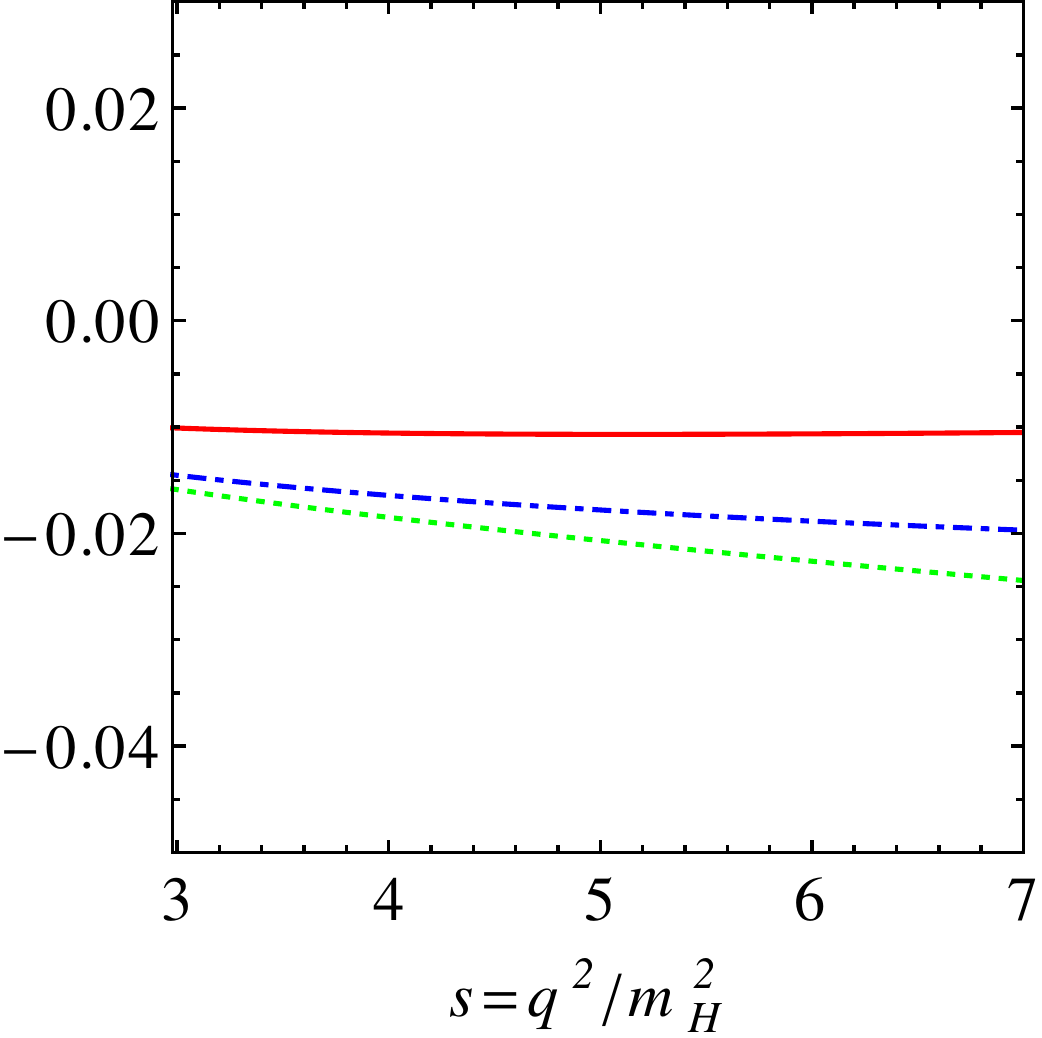}\label{Fig2b}}
\caption{(a) $d\Gamma/ds$, (b)  $-\AsymThree$.
Three scenarios with the same $\wh\alpha^V_{\Phi \ell}$ coupling are 
considered. The red solid-line is the SM case. The  dotted green line 
corresponds to $(\wh\alpha^V_{\Phi \ell},\wh\alpha^A_{\Phi \ell})=
(5,5)\times 10^{-3}$, whereas the dot-dashed blue line to  
$(\wh\alpha^V_{\Phi \ell},\wh\alpha^A_{\Phi \ell})=(5,-5)\times 10^{-3}$.}\vspace{-0.8cm}
\label{fig:IdealMom}
\end{center}     
\end{figure}

Another case of interest is the impact of the contact $HZ\ell\ell$ interactions in the cross section and asymmetries of $\eeHZ$.
Fig.~\ref{Fig2a} shows that the total cross section is rather sensitive to the axial contact coupling. Such a sensitivity is
a consequence of the higher values of $s$ probed in the scattering and is not observed in the decay rate of $\HZll$.
The asymmetry of Fig.~\ref{Fig2b} can reach a few percent and is mainly sensitive to the vector contact coupling $\alpha^V_{\Phi \ell}$.
We remark that the pattern of contributions to the different asymmetries can be well understood with the help of approximated analytic  expressions
given in Ref.~\cite{BBW14}.

\section{Conclusions}

\begin{itemize}
\item We identify several angular asymmetries, which are indeed 
very sensitive to anomalous couplings.
\item Within the presently allowed range of the anomalous 
$HZ\gamma$ interaction strength,  $\ha_{AZ}$, 
modifications of angular asymmetries 
of ${\cal O}(1)$ and even larger relative to the SM value 
are still possible indicating sensitivity to multi-TeV scales.
\item Anomalous $HZ\ell\ell$ contact interactions have smaller 
effects. This is  mainly because we find that their size is already 
tightly constrained by existing data, in agreement with 
the constraints derived in Ref.~\cite{PR13}. The effects of the contact 
$HZ\ell \ell$ interactions in the angular asymmetries of $\HZll$ 
were previously investigated in Ref.~\cite{BCD13}. While we 
formally agree with their results, we find significantly smaller 
asymmetries, since the typical values of 
$\haV$ adopted in that paper are about a factor of four 
larger than those allowed in the present analysis.
\item At present, the CP-odd $d=6$ couplings are not strongly 
constrained by data. We showed, in Ref.~\cite{BBW14}, that CP-odd asymmetry 
$\mathcal{A}_\phi^{(1)}$ can reach 
the few percent level in both in $\HZll$ decay and $\eeHZ$ 
Higgs production. In $\HZll$ an asymmetry-zero may
occur. However, for allowed values of the CP-odd couplings 
the asymmetry that can display this zero is never large.
\item Most interesting asymmetries are small in absolute 
terms, reaching at most 10\%, 
and often much less, because they are suppressed by the small 
vector $Z\ell\ell$ coupling.
\item Overall, the process $\eeHZ$ seems better suited 
than $\HZll$ for the study of anomalous $HZ\ell\ell$ 
contact interactions due to 
the higher di-lepton invariant masses.  This is particularly 
true for the contributions of $\haA$ (as well as of $\ha_{ZZ}$) 
to the total cross section, where 
$15$\% percent modifications are possible, which is illustrated in Fig.~\ref{Fig2a}. On the other hand, 
$\HZll$ provides better sensitivity to the anomalous $HZ\gamma$ 
coupling due to the photon-pole enhancement, as seen in Fig.~\ref{Fig1a}.
\end{itemize}

In Ref.~\cite{BBW14} we provided an estimate of SM loop effects, which 
 suggests that the loop effects are small compared to the present bounds on $d=6$ contributions.
Obviously, a realistic
extraction of $d=6$ couplings from high-statistics data requires the inclusion of the SM loop contributions, which were computed in the past~\cite{Kniehl90,Kniehl91,Kniehl94,BDDW06}.

Finally, our results show that the
experimental detection of angular asymmetries will be challenging
even with the planned higher statistics up-grades of the LHC.

\section*{Acknowledgements}
This work was supported in part by the Gottfried Wilhelm Leibniz program 
of the Deutsche Forschungsgemeinschaft (DFG). The work of DB was supported in part by the Alexander von Humboldt
Foundation, and the  S\~ao Paulo Research Foundation (FAPESP) grant 14/50683-0.  The work of YMW was supported by the DFG Sonderforschungsbereich/Transregio 9 ``Computergest\"utzte Theoretische Teilchenphysik".



%% References

\nocite{*}
\bibliographystyle{elsarticle-num}
%\bibliography{martin}

\begin{thebibliography}{99}

%% Discovery of the Higgs boson
\bibitem{Atlas_Higgs}
  G.~Aad {\it et al.}  [ATLAS Collaboration],
  %``Observation of a new particle in the search for the Standard Model Higgs boson with the ATLAS detector at the LHC,''
  Phys.\ Lett.\ B {\bf 716} (2012) 1,
  arXiv:1207.7214 [hep-ex].
  %%CITATION = ARXIV:1207.7214;%%

\bibitem{CMS_Higgs}
  S.~Chatrchyan {\it et al.}  [CMS Collaboration],
  %``Observation of a new boson at a mass of 125 GeV with the CMS experiment at the LHC,''
  Phys.\ Lett.\ B {\bf 716} (2012) 30,
  arXiv:1207.7235 [hep-ex].
  %%CITATION = ARXIV:1207.7235;%%

\bibitem{Atlas_HiggsQN}
  G.~Aad {\it et al.}  [ATLAS Collaboration],
  %``Evidence for the spin-0 nature of the Higgs boson using ATLAS data,''
  Phys.\ Lett.\ B {\bf 726} (2013) 120,
  arXiv:1307.1432 [hep-ex].
  %%CITATION = ARXIV:1307.1432;%%

\bibitem{CMS_HiggsQN}
S.~Chatrchyan {\it et al.}  [CMS Collaboration],
  %``Measurement of the properties of a Higgs boson in the four-lepton final state,''
  Phys.\ Rev.\ D {\bf 89} (2014) 092007,
  arXiv:1312.5353 [hep-ex].
  %%CITATION = ARXIV:1312.5353;%%
  %44 citations counted in INSPIRE as of 26 May 2014



\bibitem{BW86}
  W.~Buchm\"uller and D.~Wyler,
  %``Effective Lagrangian Analysis of New Interactions and Flavor Conservation,''
  Nucl.\ Phys.\ B {\bf 268} (1986) 621.
  %%CITATION = NUPHA,B268,621;%%

\bibitem{GIMR10}
  B.~Grzadkowski, M.~Iskrzynski, M.~Misiak and J.~Rosiek,
  %``Dimension-Six Terms in the Standard Model Lagrangian,''
  JHEP {\bf 1010} (2010) 085,
  arXiv:1008.4884 [hep-ph].
  %%CITATION = ARXIV:1008.4884;%%



%
\bibitem{CMMZ2003}
  S.~Y.~Choi, D.~J.~Miller, M.~M.~M\"uhlleitner and P.~M.~Zerwas,
  %``Identifying the Higgs spin and parity in decays to Z pairs,''
  Phys.\ Lett.\ B {\bf 553} (2003) 61
  [hep-ph/0210077].
  %%CITATION = HEP-PH/0210077;%%

\bibitem{GMM07}
R.~M.~Godbole, D.~J.~Miller and M.~M.~M\"uhlleitner,
  %``Aspects of CP violation in the H ZZ coupling at the LHC,''
  JHEP {\bf 0712} (2007) 031, 
  arXiv:0708.0458 [hep-ph].
  %%CITATION = ARXIV:0708.0458;%%
  
  
  %%%%%%% ............. to here 
  

\bibitem{DeRujula_etal10}
  A.~De Rujula, J.~Lykken, M.~Pierini, C.~Rogan and M.~Spiropulu,
  %``Higgs look-alikes at the LHC,''
  Phys.\ Rev.\ D {\bf 82} (2010) 013003,
  arXiv:1001.5300 [hep-ph].
  %%CITATION = ARXIV:1001.5300;%%

\bibitem{Bea12}
  S.~Bolognesi, Y.~Gao, A.~V.~Gritsan, K.~Melnikov, M.~Schulze, N.~V.~Tran and A.~Whitbeck,
  %``On the spin and parity of a single-produced resonance at the LHC,''
  Phys.\ Rev.\ D {\bf 86} (2012) 095031,
  arXiv:1208.4018 [hep-ph].
  %%CITATION = ARXIV:1208.4018;%%

\bibitem{IMT13}
G.~Isidori, A.~V.~Manohar and M.~Trott,
  %``Probing the nature of the Higgs-like Boson via $h \to V \mathcal{F}$ decays,''
  Phys.\ Lett.\ B {\bf 728} (2014) 131,
  arXiv:1305.0663 [hep-ph].
  %%CITATION = ARXIV:1305.0663;%%
  %19 citations counted in INSPIRE as of 26 May 2014

\bibitem{IT13}
G.~Isidori and M.~Trott,
  %``Higgs form factors in Associated Production,''
  JHEP {\bf 1402} (2014) 082,
  arXiv:1307.4051 [hep-ph].
  %%CITATION = ARXIV:1307.4051;%%
  %15 citations counted in INSPIRE as of 26 May 2014
  
  
  

 

\bibitem{GMP13}
  B.~Grinstein, C.~W.~Murphy and D.~Pirtskhalava,
  %``Searching for New Physics in the Three-Body Decays of the Higgs-like Particle,''
  JHEP {\bf 1310} (2013) 077,
  arXiv:1305.6938 [hep-ph].
  %%CITATION = ARXIV:1305.6938;%%


\bibitem{GI14}
M.~Gonzalez-Alonso and G.~Isidori,
  %``The $h \to 4 \ell$ spectrum at low $m_{34}$: Standard Model vs. light New Physics,''
  arXiv:1403.2648 [hep-ph].
  %%CITATION = ARXIV:1403.2648;%%


\bibitem{BCD13}  G.~Buchalla, O.~Cat\`a and G.~D'Ambrosio,
  %``Nonstandard Higgs Couplings from Angular Distributions in $h\to Z \ell^+\ell^-$,''
Eur.~Phys.~J. C {\bf 74} (2014) 2798, arXiv:1310.2574 [hep-ph].
  %%CITATION = ARXIV:1310.2574;%%

 
\bibitem{BBW14}  M.~Beneke, D.~Boito and Y.~M.~Wang,
  %``Anomalous Higgs couplings in angular asymmetries of H --> Zl+l- and e+e- --> HZ,''
  JHEP {\bf 1411} (2014) 028, arXiv:1406.1361 [hep-ph]
  %%CITATION = ARXIV:1406.1361;%%
  %6 citations counted in INSPIRE as of 04 Nov 2014



\bibitem{Baeretal13}
  H.~Baer {\it et al.},
  %``The International Linear Collider Technical Design Report - Volume 2: Physics,''
  arXiv:1306.6352 [hep-ph].
  %%CITATION = ARXIV:1306.6352;%%
  %59 citations counted in INSPIRE as of 25 Feb 2014


%%%%%%%%%%%%%%%%%%%%%%%%%%


%% References to e+e- ---> H Z

\bibitem{Barger:1993wt}
V.~D.~Barger, K.~M.~Cheung, A.~Djouadi, B.~A.~Kniehl and P.~M.~Zerwas,
  %``Higgs bosons: Intermediate mass range at e+ e- colliders,''
  Phys.\ Rev.\ D {\bf 49} (1994) 79
  [hep-ph/9306270].
  %%CITATION = HEP-PH/9306270;%%
  %205 citations counted in INSPIRE as of 04 Jun 2014


\bibitem{Hagiwara:1993sw}
  K.~Hagiwara and M.~L.~Stong,
  %``Probing the scalar sector in e+ e- ---> f anti-f H,''
  Z.\ Phys.\ C {\bf 62} (1994) 99
  [hep-ph/9309248].
  %%CITATION = HEP-PH/9309248;%%
  %104 citations counted in INSPIRE as of 04 Jun 2014

\bibitem{KKZ96}
  W.~Kilian, M.~Kr\"amer and P.~M.~Zerwas,
  %``Anomalous couplings in the Higgsstrahlung process,''
  Phys.\ Lett.\ B {\bf 381} (1996) 243
  [hep-ph/9603409].
  %%CITATION = HEP-PH/9603409;%%


%
\bibitem{Heinemeyer_etal13}
  S.~Heinemeyer {\it et al.}  [The LHC Higgs Cross Section Working Group Collaboration],
  %``Handbook of LHC Higgs Cross Sections: 3. Higgs Properties,''
  arXiv:1307.1347 [hep-ph].
%


\bibitem{AJMT13}
R.~Alonso, E.~E.~Jenkins, A.~V.~Manohar and M.~Trott,
  %``Renormalization Group Evolution of the Standard Model Dimension Six Operators III: Gauge Coupling Dependence and Phenomenology,''
  JHEP {\bf 1404} (2014) 159,
  arXiv:1312.2014 [hep-ph].
  %%CITATION = ARXIV:1312.2014;%%
  %16 citations counted in INSPIRE as of 26 May 2014

%
  
\bibitem{PR13}A.~Pomarol and F.~Riva,
  %``Towards the Ultimate SM Fit to Close in on Higgs Physics,''
  JHEP {\bf 1401} (2014) 151,
  arXiv:1308.2803 [hep-ph].
  %%CITATION = ARXIV:1308.2803;%%
  %28 citations counted in INSPIRE as of 26 May 2014

\bibitem{Kniehl90}
  B.~A.~Kniehl,
  %``Radiative corrections for $H \to Z Z$ in the standard model,''
  Nucl.\ Phys.\ B {\bf 352} (1991) 1.
  %%CITATION = NUPHA,B352,1;%%

%\cite{Kniehl:1991hk}
\bibitem{Kniehl91}
  B.~A.~Kniehl,
  %``Radiative corrections for associated $Z H$ production at future $e^{+} e^{-}$ colliders,''
  Z.\ Phys.\ C {\bf 55} (1992) 605.
  %%CITATION = ZEPYA,C55,605;%%

\bibitem{Kniehl94}
    B.~A.~Kniehl,
  %``Higgs phenomenology at one loop in the standard model,''
  Phys.\ Rept.\  {\bf 240} (1994) 211.
  %%CITATION = PRPLC,240,211;%%

%%%%%%%
%%% SM loops
\bibitem{BDDW06}
A.~Bredenstein, A.~Denner, S.~Dittmaier and M.~M.~Weber,
  %``Precise predictions for the Higgs-boson decay H ---> WW/ZZ ---> 4 leptons,''
  Phys.\ Rev.\ D {\bf 74} (2006) 013004
  [hep-ph/0604011].
  %%CITATION = HEP-PH/0604011;%%




%%%%%%%%%%


\end{thebibliography}

\end{document}